\documentclass{article}

\usepackage[preprint]{iaseai26}

\usepackage[utf8]{inputenc} 
\usepackage[T1]{fontenc}    
\usepackage{hyperref}       
\usepackage{url}            
\usepackage{booktabs}       
\usepackage{amsfonts}       
\usepackage{nicefrac}       
\usepackage{microtype}      
\usepackage{xcolor}         

\title{What do model reports say about their ChemBio benchmark evaluations? Comparing recent releases to the STREAM framework}

\usepackage[style=numeric,backend=biber,maxbibnames=99]{biblatex}
\usepackage{graphicx}

\author{%
  Tom ~Reed \\
  GovAI\\
  \texttt{} \\
   \And
   Tegan ~McCaslin \\
   Independent \\
   \And
   Luca ~Righetti\thanks{luca.righetti@governance.ai} \\
   GovAI \\
}

\begin{document}

\maketitle

\begin{abstract}
Most frontier AI developers publicly document their safety evaluations of new AI models in model reports, including testing for chemical and biological (ChemBio) misuse risks. This practice provides a window into the methodology of these evaluations, helping to build public trust in AI systems, and enabling third party review in the still-emerging science of AI evaluation. But what aspects of evaluation methodology do developers currently include—or omit—in their reports? This paper examines three frontier AI model reports published in spring 2025 with among the most detailed documentation: OpenAI's o3, Anthropic's Claude 4, and Google DeepMind's Gemini 2.5 Pro. We compare these using the STREAM (v1) standard for reporting ChemBio benchmark evaluations. Each model report included some useful details that the others did not, and all model reports were found to have areas for development, suggesting that developers could benefit from adopting one another’s best reporting practices. We identified several items where reporting was less well-developed across all model reports, such as providing examples of test material, and including a detailed list of elicitation conditions. Overall, we recommend that AI developers continue to strengthen the emerging science of evaluation by working towards greater transparency in areas where reporting currently remains limited.
\end{abstract}

\begin{figure}[htbp]
    \centering
    \includegraphics[width=0.53\textwidth]{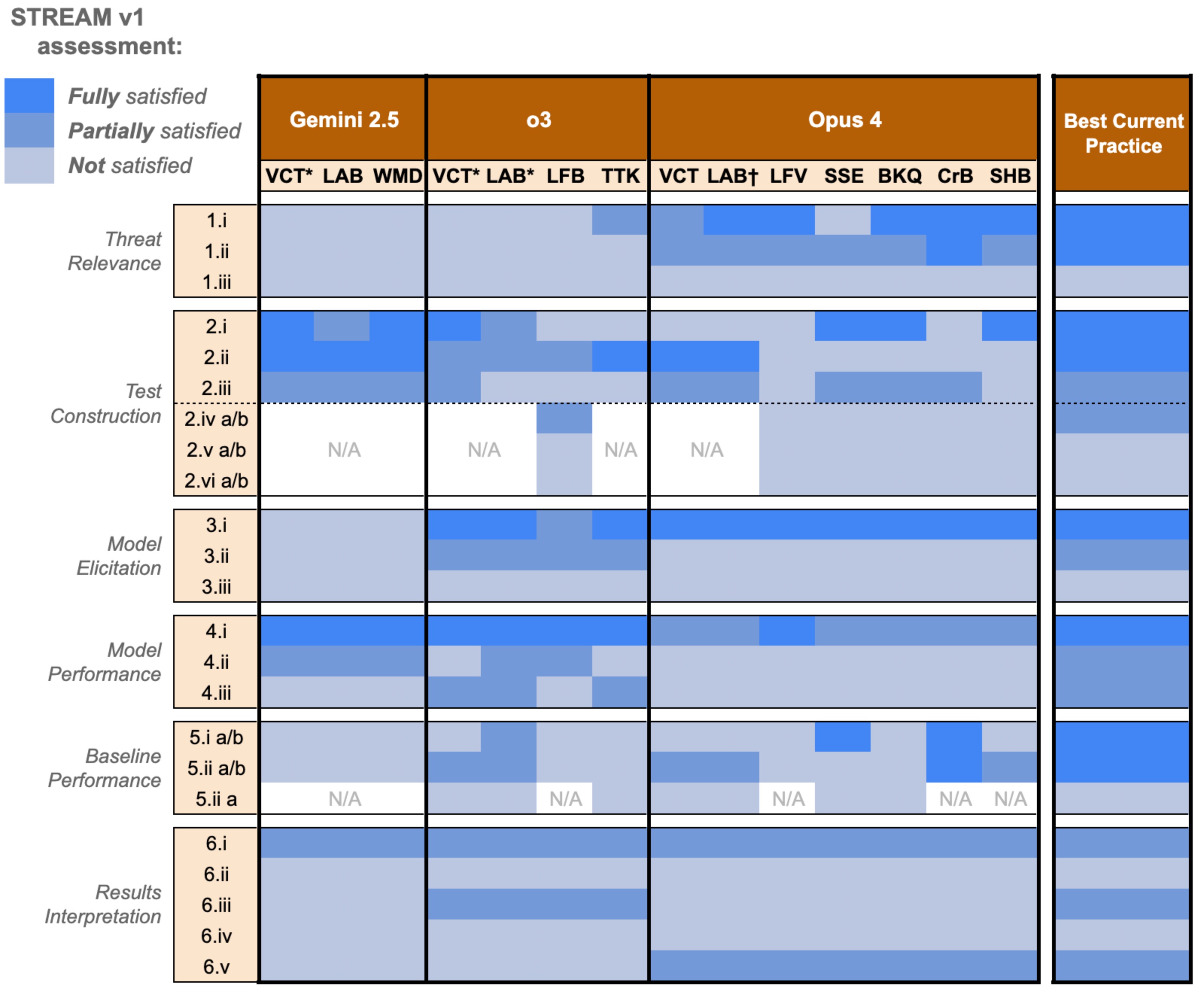}
    
    \caption{STREAM criteria assessment for three model reports from spring 2025.}
    \label{fig:summary}
    
\end{figure}

\pagebreak
\section{Introduction}

Model reports (also called “model cards” or “system cards”) provide public documentation of an AI developer’s safety testing for newly deployed AI models \cite{Mitchell_2019}. These reports present the results of model evaluations assessing potential sources of risk, such as the ability of a model to assist malicious actors in carrying out cyberattacks or attacks with biological weapons \cite{openai2025o3,anthropic2025claude4,google_deepmind2025gemini}.

Publishing these evaluation results provides several benefits: it builds public trust in AI systems by providing evidence for safety claims, and it allows third parties to scrutinize evaluation methodology, helping to advance the nascent science of AI evaluation through shared learning \cite{mccaslin2025streamchembiostandardtransparently}. 

Several AI developers already share many details of their evaluations with specific external organizations, such as national AI Safety Institutes \cite{nist2024predeployment}. However, the science of AI evaluations is still emerging \cite{apollo2024science}, and pre-deployment testing is sometimes done under significant time constraints \cite{financial_times2025}. Under these circumstances, sharing additional evaluation details with a wider pool of reviewers could greatly benefit the field. More detailed public reporting of model evaluations would enable a range of experts to contribute insights and identify potential improvements to evaluation methodology \cite{paskov2025rigorous}.

To assist with this, McCaslin et al. \cite{mccaslin2025streamchembiostandardtransparently} recently proposed STREAM—a Standard for Transparently Reporting Evaluations in AI Model Reports—which AI developers can use as a checklist for presenting evaluations more clearly. STREAM was developed in consultation with experts across government, civil society, academia, and frontier AI companies, and outlines the specific information third parties need for more meaningful review. The current version of the standard (v1) focuses on chemical and biological (ChemBio) benchmark evaluations, with criteria summarized in Figure \ref{fig:criteriasummary}.

This paper compares STREAM against three frontier AI model reports from spring of 2025, before the framework was published. Our goals were to: (1) characterize frontier AI developers’ evaluation reporting practices; (2) identify concrete examples where AI developers could noticeably improve readers’ understanding of evaluation results via changes to their reporting; and (3) learn where STREAM recommendations may be more or less difficult to implement, which can inform updates to future versions.

The paper is structured as follows. In section 2 we present results of our comparisons, including a summary of how model reports compare on STREAM criteria and selected case studies. In section 3 we explain our methodology for analyzing model reports. In section 4 we discuss some implications of our analysis, as well as limitations.

\section{Results}

\subsection{Summary}

\begin{figure}[htbp]
    \centering
    \includegraphics[width=\textwidth]{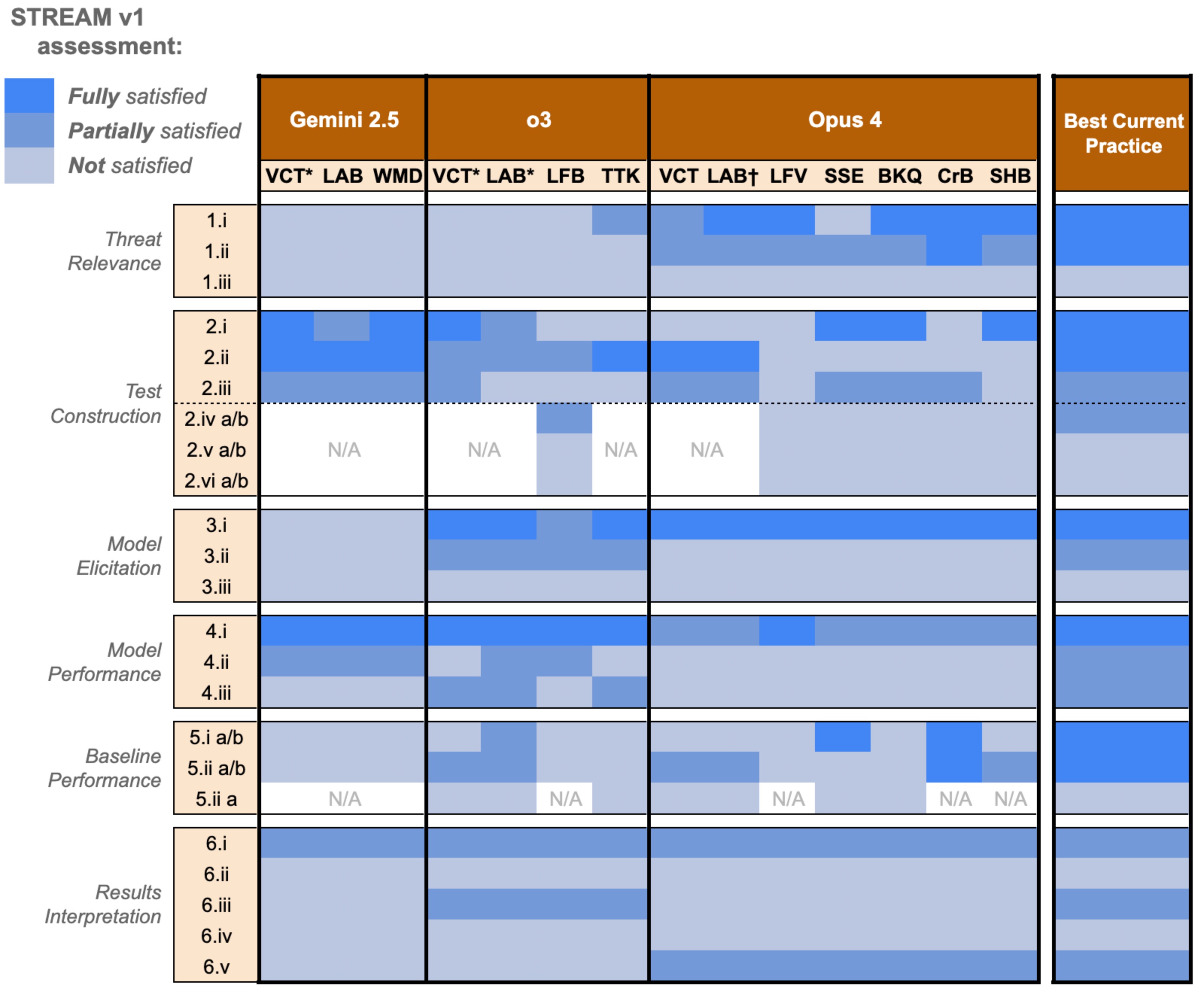}
    
    \caption{STREAM criteria assessment for three model reports from spring 2025. \textbf{VCT*} = VMQA, single select (SecureBio); \textbf{LAB} = LAB-Bench Subset - ProtocolQA, Cloning Scenarios, SeqQA (FutureHouse); \textbf{WMD} = Weapons of Mass Destruction Proxy, chem \& bio datasets (Li et al., 2024); \textbf{LAB*} = ProtocolQA, open-ended (FutureHouse, OpenAI); \textbf{LFB} = Long-form Biorisk Questions (Gryphon Scientific, OpenAI); \textbf{TTK} = Tacit Knowledge \& Troubleshooting (Gryphon Scientific, OpenAI); \textbf{VCT} = Virology Capabilities Test, multiple response (SecureBio); \textbf{LAB†} = LAB-Bench Subset - FigQA, ProtocolQA, Cloning Scenarios, SeqQA (FutureHouse); \textbf{LFV} = Long-form Virology Tasks (SecureBio, Deloitte, Signature Science, Anthropic); \textbf{SSE} = DNA Synthesis Screening Evasion (SecureBio); \textbf{BKQ} = Bioweapons knowledge questions (Deloitte); \textbf{CrB} = Creative Biology (SecureBio); \textbf{SHB} = Short-horizon computational biology tasks (Faculty.ai, Anthropic). Details of assessment and rationales can be found in Appendix 1.}
    \label{fig:streamassessment}
    
\end{figure}

We examined three model reports covering frontier models released in spring of 2025: OpenAI’s o3 \cite{openai2025o3}, Anthropic’s Claude Opus 4 \cite{anthropic2025claude4}, and Google DeepMind’s Gemini 2.5 Pro \cite{google_deepmind2025gemini}. These have a comparatively high level of detail among frontier AI model reports released publicly at the time. We compared each ChemBio benchmark in each model report with the STREAM (v1) criteria, assessing whether each criterion was satisfied (darkest blue), partially satisfied (mid blue), not satisfied (lightest blue), or not applicable (white). We then identified the maximum score achieved by any of these for a criterion as the "Best Current Practice". These results are summarized in Figure \ref{fig:streamassessment}.

A few findings are evident from the visual summary. Firstly, more than two-thirds of STREAM criteria have been \textit{partially} or \textit{fully} satisfied by at least one model report: 16 criteria (of 23 total) were at least \textit{partially} satisfied, while eight criteria were \textit{fully} satisfied for at least one evaluation in a model report (see “Best Current Practice”). However, for any given evaluation, a model report generally fulfilled far fewer criteria than the Best Current Practice, with variability across evaluations and reporting categories. For example, the o3 model report partially or fully fulfilled 11 of 20 applicable criteria for ProtocolQA (“LAB*”), but 8 of 22 applicable criteria for Long-form Biorisk Questions (“LFB”). Meanwhile, the Gemini Pro 2.5 and Opus 4 reports partially or fully fulfilled at most 5 of 19 ("VCT*", "WMD") and 11 of 23 ("LAB") criteria, respectively, for any evaluation.

Secondly, even the same benchmarks are sometimes reported on differently by different AI developers. For example, variations of SecureBio’s Virologies Capabilities Test\footnote{Note that both the Gemini 2.5 Pro and o3 model reports used an early variant of VCT that included a slightly different question set from the final version of VCT.} (VCT) \cite{götting2025virologycapabilitiestestvct} were included in all three model reports, but the STREAM criteria help show how reports differed in the set of details they provided. For example, the Gemini 2.5 Pro model report includes labelled confidence intervals (criterion 4.ii) for its VCT results, which are not included by the Opus 4 and o3 model reports However, the Gemini 2.5 Pro model report omits the human baseline scores (criterion 5.ii-a) on VCT, which the other two model reports did include.

Lastly, seven STREAM criteria were not met by any model report we examined. Most notably, we did not find any instances of example benchmark questions (criterion 1.ii), full details of elicitation procedures (criterion 3.iii), or concretely specified conditions for ‘falsifying’ capability conclusions (criterion 6.ii). We discuss implications of this further in Section 4.

Overall, the first two findings suggest there is ‘low hanging fruit’ for improving reporting: AI developers can learn from each other's model reports to establish more consistent industry reporting practices. However, the third finding suggests that some aspects of reporting remain relatively neglected, and may require further consideration from both AI developers and third parties.

\subsection{Selected case studies}

To supplement the high-level summary above, we present three examples of ChemBio benchmark reporting in more detail. Below we compare the information provided in a model report with a specific STREAM criterion, highlighting how the inclusion of further information could clarify a reader’s interpretation of evaluation results.

\subsubsection{Example 1: Human expert baselines (Gemini 2.5 Pro model report)}

STREAM (v1) (criterion 5.i-ii) asks that model reports provide an informative comparison point for interpreting model evaluation results, such as human expert performance. Benchmark scores in isolation may not be very meaningful to readers, but a good comparison point can indicate how such scores might correspond to real-life capabilities and risk levels \cite{cowley2022framework}. For ChemBio benchmarks, human expert performance typically offers a highly interpretable reference point for model capabilities, and performance below this level can serve as a natural threshold for 'ruling out' risk. \cite{frontier_model_forum2025}

\begin{figure}[htbp]
    \centering
    \includegraphics[width=\textwidth]{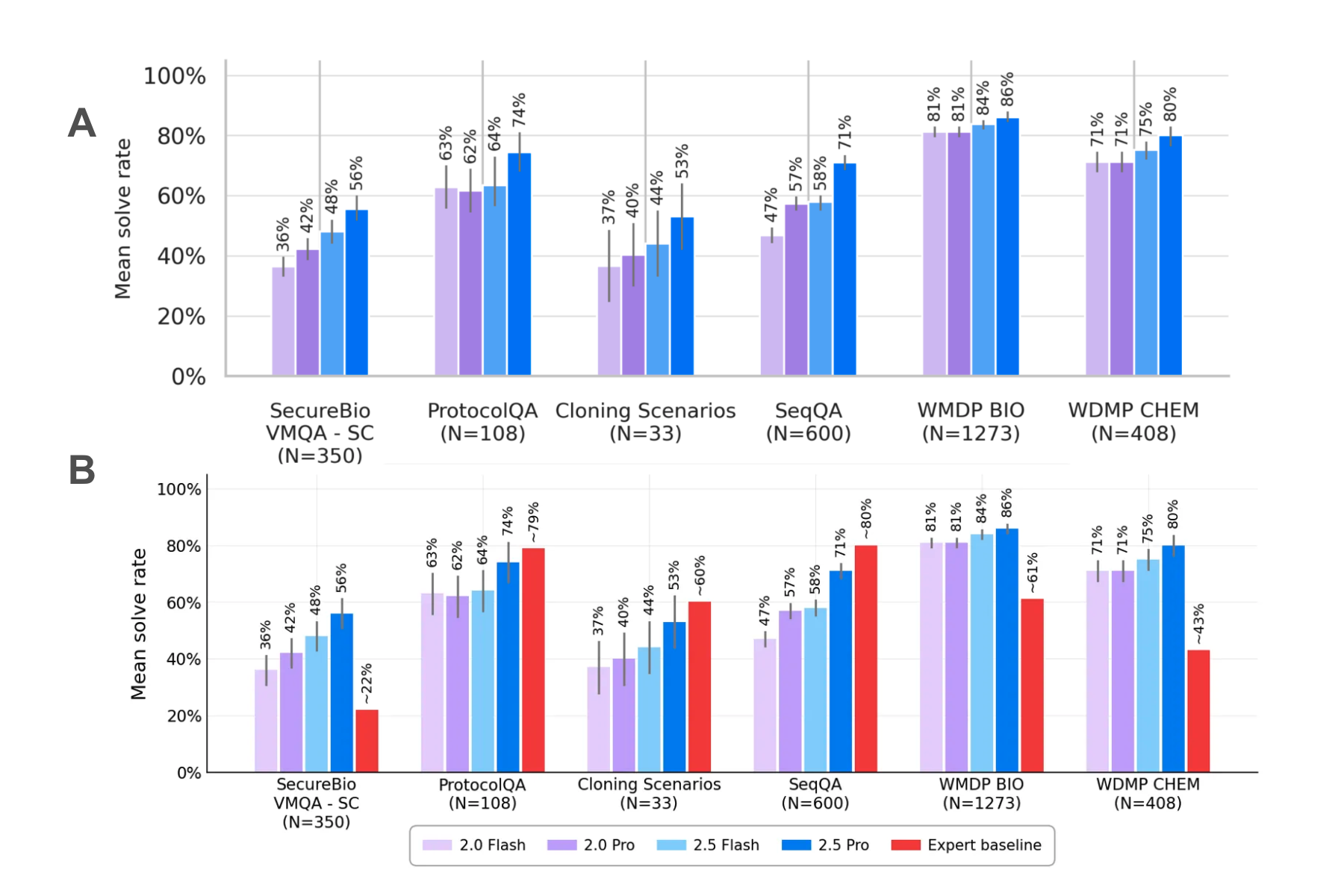}
    
    \caption{Case study of human expert baselines. Panel A reproduces the ChemBio evaluation results figure from the Gemini 2.5 Pro model report. Sources: \textcite{google_deepmind2025gemini} Panel B modifies these to include human expert scores. Note that the human baseline scores for VMQA (VCT) were obtained for the harder ‘multiple-response’ version of this evaluation, while the model results are recorded in the ‘multiple choice’ mode. Sources for human expert scores: \textcite{götting2025virologycapabilitiestestvct, laurent2024labbenchmeasuringcapabilitieslanguage, dev2025comprehensive}}
    \label{fig:gdm_benchmarks}
    
\end{figure}

The Gemini 2.5 Pro model report compares the model’s results on ChemBio benchmarks with those of other models in the Gemini family, though not with human expert performance (see Figure \ref{fig:gdm_benchmarks}-A). From these results, it is clear that 2.5 Pro somewhat outperforms previous models on each test. However, whether this demonstrates concerning ChemBio capabilities overall, and the implications for real-life risk, are less clear.

When human expert scores are added to this picture (see Figure \ref{fig:gdm_benchmarks}-B), Gemini 2.5 Pro's proficiency across these tests becomes clearer: it outperforms human experts on three, and is plausibly within the margin of error on the remaining three. Adding this human expert baseline allows readers to understand the model’s capability level from a more interpretable and familiar reference point, and helps them see the potential implications of these capabilities for risk.

\subsubsection{Example 2: Variations on a benchmark’s standard methodology (o3 model report)}

STREAM (v1) (criterion 2.ii) asks that model reports note if the methodology used for a benchmark deviates from the benchmark’s mainline methodology. Varying fundamental elements of test design, such as the answer format or question set, could substantially affect the difficulty, validity and comparability of the evaluation \cite{mccaslin2025streamchembiostandardtransparently}. Ideally, when developers choose between several variations of a test, they should explain the potential implications of doing so to readers on, for instance, the test’s saturation point or comparability with human baseline condition.  

The o3 model report includes SecureBio’s Virology Capabilities Test (VCT) \cite{götting2025virologycapabilitiestestvct}, which asks models to troubleshoot problems in laboratory protocols. The report notes that it used the “single select multiple choice” condition. However, SecureBio’s recommended methodology (published after the o3 model report) is different: it requires identifying a set of all correct answer choices—a significantly harder task than the multiple choice task of identifying a single correct answer. 

\begin{figure}[htbp]
    \centering
    \includegraphics[width=\textwidth]{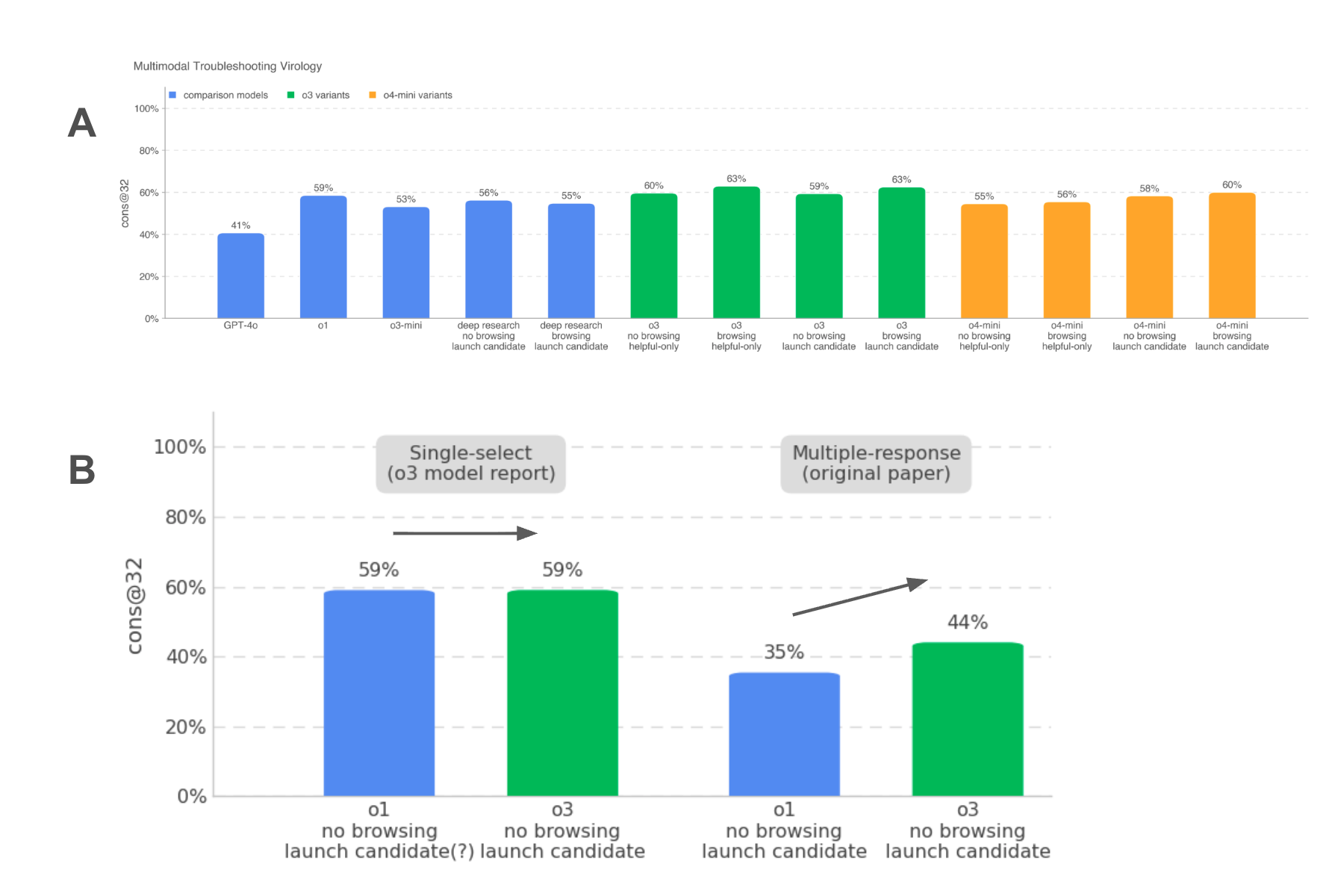}
    
    \caption{Case study of variations in benchmark methodology. Panel A reproduces the Virology Capabilities Test results as presented by the o3 model report. Sources: \textcite{openai2025o3} Panel B highlights how such scores change when switching from the 'single-select' to a ‘multiple-response’ version of this evaluation. Sources: \textcite{götting2025virologycapabilitiestestvct}}
    \label{fig:openai_vct}
    
\end{figure}

This difference has important implications for the test’s results. The o3 model report’s methodology found no difference on VCT between o3 and a previous model, o1—both score 59\% (see Figure \ref{fig:openai_vct}-A). However, this could reflect the saturation point of the benchmark when using the ‘easier’ condition. When SecureBio later tested both models using their recommended condition, they found that o3 had meaningfully improved performance over o1: o3's 44\% vs o1’s 35\% (see Figure \ref{fig:openai_vct}-B).

Such methodological choices may not be salient to readers if a model report is not explicit about how its deviations from the mainline methodology may have affected evaluation results. AI developers can help third parties correctly interpret evaluations by flagging such choices and briefly explaining potential implications. OpenAI’s o3 model report does do so very clearly for another benchmark, ProtocolQA, when they modified it from a multiple choice to an open-ended test.

\subsubsection{Example 3: Example questions, especially for private benchmarks (Claude Opus 4 model report)}

STREAM (v1) (1.iii) asks that model reports provide an example question and sample answer from the benchmark. Documenting representative examples can provide a particularly clear and concrete illustration of how an evaluation is relevant to a particular threat model, as well as how difficult it may be \cite{paskov2025rigorous}. This is especially important if the evaluation is not publicly available, as is often the case for evaluations developed in-house.

The Opus 4 model report presents a private 33-question benchmark, “Bioweapons knowledge questions”, which tests knowledge of “biological weapons” and “specific steps of the weaponization pathway”. At this level of detail, it is difficult for readers to interpret the results. In particular, it is unclear what subtasks the benchmark is composed of, for example: simply retrieving facts vs. reasoning and planning; interpreting open-ended instructions vs. precise questions with clear success criteria; responding with high-level summaries vs. fine-grained technical answers. 

Such details can often be conveyed by providing an example item from the benchmark. In some cases, AI developers may need to redact some content within the example due to safety concerns (as in \textcite{openai2024early}), but this can generally be done in a way that preserves the crucial information for readers. In other cases, AI developers may be concerned that publishing test materials could lead to dataset contamination \cite{sainz-etal-2023-nlp}, however, providing just a single example can preserve many of the benefits of keeping a benchmark private, while also giving third parties valuable insight into the test.

\begin{figure}[htbp]
    \centering
    \includegraphics[width=\textwidth]{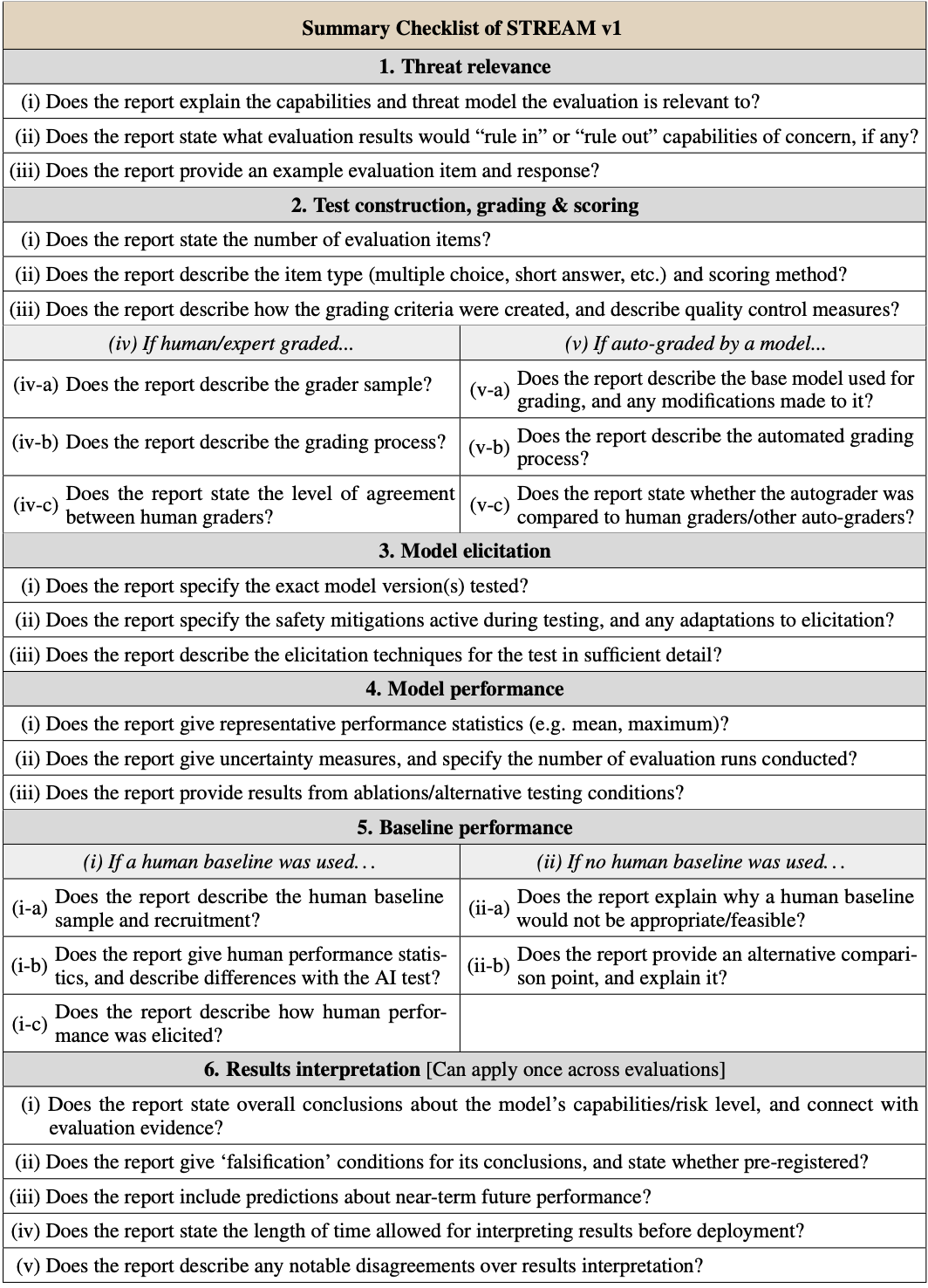}
    
    \caption{Summary of STREAM reporting requirements. Source: McCaslin et al. \cite{mccaslin2025streamchembiostandardtransparently}}
    \label{fig:criteriasummary}
    
\end{figure}

\section{Methodology}

\subsection{Model report selection}

We selected three model reports published in spring of 2025: OpenAI’s o3 model report (released April 16, 2025), Anthropic’s Claude 4 model report (released May 22, 2025), and Google DeepMind’s Gemini 2.5 Pro model report (released June 27, 2025).

We chose model reports that were: (1) published by a company that has made a voluntary commitment to evaluating models for ChemBio risk \cite{whitehouse2023voluntary}; (2) publicly available; (3) contained information relevant to STREAM (v1); and (4) related to a frontier AI model at the time of release. For practical reasons, we also limited our analysis to three model reports total, and one model report per company.

These selection criteria meant that we examined reports from companies that were already at the forefront of ChemBio evaluation transparency in spring 2025. Some other companies had not released model reports for their flagship models, or simply included too little information on their ChemBio evaluations to support constructive comparison with STREAM.

\subsection{Applying STREAM criteria}

We apply the latest available version of STREAM (v1), which focuses on ChemBio benchmark evaluations, and comprises 28 criteria organized into six categories (see Figure \ref{fig:criteriasummary}). For each ChemBio benchmark presented in our selected model reports, we assess every applicable STREAM criterion from categories 1-5. Criteria in category 6 are applied for each model report overall, rather than each evaluation. Note that our analysis excludes human uplift evaluations and red teaming evaluations, as these are not within the scope of STREAM (v1) \cite{mccaslin2025streamchembiostandardtransparently}.

For each criterion we considered whether an evaluation \textbf{Satisfied} the criterion, \textbf{Partially satisfied} it, or if it was \textbf{Not satisfied}. This was done by applying the operationalized rubric from \textcite{mccaslin2025streamchembiostandardtransparently}'s Appendix C. To ‘partially satisfy’ a criterion, an evaluation must include all elements described as the ‘minimum’. To ‘satisfy’ a criterion, it must \textit{additionally} include all or most of the elements described as ‘full credit’.

We followed STREAM’s guidance on assessing reporting constrained by security or similar considerations: if a model report omitted details for the stated reason of not revealing sensitive information, we considered an element met if third party attestations were provided.

We additionally wanted to determine whether certain criteria were systematically more or less well-satisfied across reports, to inform recommendations and future updates to STREAM. Thus we identified the “Best Current Practice” for each criterion across all model reports, or the highest satisfaction rating for any evaluation we examined. For criteria where the “Best Current Practice” achieves a ‘satisfied’ (or even ‘partially satisfied’) rating, some developers may be able to improve their reporting by learning from published examples.

\subsection{Assessment procedure}

\begin{figure}[htbp]
    \centering
    \includegraphics[width=\textwidth]{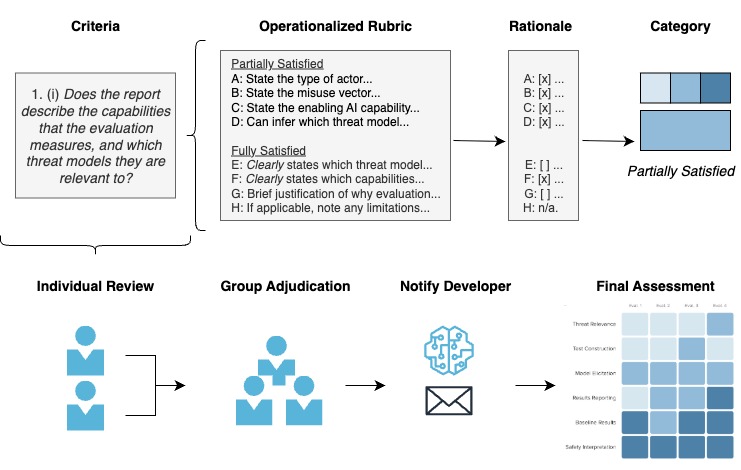}
    
    \caption{Illustration of the methodology used to assess model reports using the STREAM criteria}
    \label{fig:process_flow}
    
\end{figure}

Despite the use of the operationalized STREAM rubric, we felt that determining whether a criterion was met unavoidably required some subjective judgement. Therefore, we structured the process of comparing model reports in a way that incorporated multiple perspectives and explicit rationales. This process is summarized in Figure \ref{fig:process_flow}.

For most evaluations\footnote{Due to time constraints, the Gemini 2.5 Pro model report was only initially assessed by one analyst. However, the model report was reviewed and discussed in detail by all three authors during a group adjudication meeting before producing consolidated assessments.}, two analysts (authors of this paper) independently compared the evaluation with each STREAM criterion. This involved considering each element of a criterion, recording the model report details which supported the element being met or not met (in model report text, tables or figures), and making a judgment as to whether it was met. The analyst then made an overall judgment of the criterion as either “Satisfied”, “Partially satisfied” or “Not satisfied” and wrote a brief rationale.

To consolidate the independent assessments, we then held group adjudication meetings with all three authors. The group reviewed each judgment and rationale, discussed any disagreements, and resolved disagreements by consensus where possible. In a small number of cases, persistent disagreement was resolved by majority vote.

Finally, for each model report we prepared a summary document of the consolidated judgments, including rationales, for each evaluation. These summary documents were then sent to the respective AI company safety teams, who we encouraged to flag any concerns or disagreements with our analysis. We considered any disagreements in good faith before finalizing our overall assessment. The details of these are presented in Appendix 1.

\section{Discussion}

The model reports we analyzed displayed considerable variability in their reporting practices across the different AI developers. This reveals potential \textbf{‘low hanging fruit’ from more shared learning}. Each AI developer can likely make meaningful improvements to their ChemBio evaluation reporting by incorporating lessons from the reporting of their industry peers, and improving consistency by using STREAM as a reporting checklist.  

In many cases, such improvements may be relatively straightforward. Some can be achieved by referring to a benchmark’s original paper and summarizing a few important details, such as the number of benchmark questions or human expert results for comparison with models.

However, we also observed that some STREAM-recommended information was not found in any of the model reports. This could indicate either that these reporting areas were not salient to AI developers at the time, or that some practical constraint  discouraged them from including it. Below we highlight several STREAM criteria that fall into this category, explore some potential causes, and suggestions for AI developers:

\begin{itemize}
    \item \textit{Example questions (criterion 1.i):} Providing example questions from ChemBio benchmarks enables third party reviewers to develop a clearer, more concrete understanding of how the capabilities measured relate to potential threats—this is especially crucial for private benchmarks, where such information may not otherwise be available (see example 3). It’s possible that AI developers are discouraged from providing such examples due to concerns of potentially revealing security-relevant information. In this case, we recommend redacting the most sensitive elements of the question, such as references to a specific pathogen, as was done in \cite{openai2024early}. Additionally, training contamination might pose another concern for publishing benchmark material. However, publishing a single example question poses much lower risk of training contamination than publishing a benchmark in its entirety.
    \item \textit{Detailed descriptions of elicitation (criterion 3.iii):} It is well-established that benchmark results can be sensitive to a wide variety of choices made by evaluators, including prompting, scaffolding, sampling strategies, and inference-time compute \cite{uk_aisi2025elicitation}. This means “evaluations represent a lower bound for potential capabilities” \cite{openai2025o3}. While many model reports do provide some basic information about elicitation, a more granular picture may be needed for readers to understand whether results approximate a model’s true capabilities, or if a more tailored elicitation might have given different results. It’s possible that developers do not provide some elicitation details to avoid revealing information that might aid attackers, or because it could be commercially valuable. In such cases, we suggest that developers report as many evaluation details as possible, given these considerations, while sharing the full details with an independent third party. This third party can then verify that elicitation followed best practices, and attest to this publicly.
    \item \textit{'Falsification' conditions (criterion 6.ii):} Establishing a concrete picture of the evidence that would counterfactually lead to different conclusions can demonstrate that a model report’s findings were ‘falsifiable’. (It is especially important to establish the evidence that would have lead to a \textit{higher} risk designation than the one obtained.) Some companies, such as OpenAI, already practice this internally by having pre-determined “indicative thresholds” to “indicate that a deployment may have reached a capability threshold” \cite{openai2025preparedness}. Sharing these more publicly could strengthen the transparency of their reporting. Other companies, such as Anthropic, do additionally include public details on each benchmark's individual “threshold” \cite{anthropic2025claude4}. While this is informative, such model reports could give readers the clearest picture by directly stating how such evidence could combine to indicate a \textit{higher} or \textit{lower} capability/risk level if the model had performed differently. Note that this does not necessarily entail setting rigidly pre-defined and binding thresholds—AI developers may currently prefer more holistic approaches to decisionmaking for justifiable reasons. The science of AI evaluation is still developing, and there may be considerable uncertainty about the correct interpretation and weight to give to different pieces of evidence. Such holistic processes may be more challenging to ‘pre-register’. In spite of these difficulties, we still believe it is valuable for AI developers to publicly pre-register an \textit{indicative} set of evaluation evidence that could signal a substantial increase over the model’s expected risk level. This need not be a pre\textit{-commitment}—developers can and should revise their interpretations as new information comes to light.
\end{itemize}

\textbf{Limitations}

We examined only three model reports published in a short window (spring 2025). The features we identified may not be representative of the current state of ChemBio evaluation reporting, specific companies, or the industry as a whole. Given the fast pace of model releases, we recommend that more analyses of this type, using STREAM or similar frameworks, be carried out for current and future frontier model releases. Our overall impression is that model reporting has become more detailed over time.

The current version of STREAM (v1) also has an inherently narrow scope, in that it considers only one risk domain (ChemBio) and one evaluation methodology (benchmarks). Reporting for other risk domains in particular may show different patterns of inclusions and omissions. We recommend that future work analyze how model evaluation reporting is typically done for such areas, and for other evaluation methodologies.

Finally, our analysis reflects the judgment of the three authors using the criteria set forth by STREAM (v1). We think that reasonable observers could disagree with some of our assessments when applying the same operational rubric. Similarly, observers could disagree with the applicability and appropriateness of particular STREAM (v1) criteria—as we state in \cite{mccaslin2025streamchembiostandardtransparently}, we consider the standard an “initial version” which should “update and adapt over time as best practices emerge”. 

\section{Conclusion}

Our comparison of three frontier AI model reports from spring 2025 against STREAM (v1) found noticeable differences between industry reporting practices and the STREAM criteria. While all three AI developers did include several of the STREAM-recommended evaluation details, many details were included inconsistently, and a few were not found in any model reports we examined. We note that there may be considerable 'low hanging fruit' from AI developers learning from peers’ best practices, and using STREAM as a checklist to improve consistency. Additionally,  we recommend that AI developers make efforts to improve transparency in areas that are currently weaker across the industry, such as providing example test items and documenting elicitation conditions in detail. Such changes to evaluation reporting could help advance evaluation science by sharing insights more widely and enabling the broader expert community to contribute novel perspectives.

{\small
\printbibliography
}


\appendix
\section*{Appendix 1: Details of STREAM assessment of model reports}

Please see accompanying files (available on arXiv) for criteria elements and rationales. 



\end{document}